\documentclass[aps,prl,preprint,superscriptaddress]{revtex4-1}
\usepackage{amsmath}
\usepackage{graphicx}
\usepackage{amssymb}
\usepackage{amsthm}
 \usepackage{dsfont}

\DeclareMathOperator{\Tr}{Tr}

\begin{document}
\title{ Entanglement, Particle Identity \\and the GNS Construction: A Unifying Approach}
\author{A.P. Balachandran}
\email[]{bal@phy.syr.edu}
%\homepage[]{Your web page}
%\thanks{}
\affiliation{Institute of Mathematical Sciences, CIT Campus, Taramani, Chennai 600113, India}
\affiliation{Physics Department, Syracuse University, Syracuse, NY, 13244-1130, USA}

\author{T.R. Govindarajan}
\email[]{trg@imsc.res.in}
%\homepage[]{Your web page}
%\thanks{}
%\altaffiliation{}
\affiliation{Institute of Mathematical Sciences, CIT Campus, Taramani, Chennai 600113, India}

\author{Amilcar R. de Queiroz}
\email[]{amilcarq@unb.br}
%\homepage[]{Your web page}
%\thanks{}
\affiliation{Instituto de Fisica, Universidade de Brasilia, Caixa Postal 04455, 70919-970, Brasilia, DF, Brazil}
\altaffiliation{Institute of Mathematical Sciences, CIT Campus, Taramani, Chennai 600113, India}

\author{A.F. Reyes-Lega}
\email[]{anreyes@uniandes.edu.co}
\homepage[]{http://fisicateorica.uniandes.edu.co/anreyes/}
%\thanks{}
\affiliation{Departamento de F\'isica, Universidad de los Andes, A.A. 4976, Bogot\'a D.C., Colombia}
\altaffiliation{Institute of Mathematical Sciences, CIT Campus, Taramani, Chennai 600113, India}

\date{\today}

\begin{abstract}
A novel approach to entanglement, based on the Gelfand-Naimark-Segal (GNS) construction, is introduced.
It considers states as well as algebras of observables on an equal footing.
The conventional approach to the emergence of mixed from pure ones  based on taking partial traces is replaced by the more general notion of the restriction of a state to a subalgebra.  For bipartite systems of nonidentical particles, this
approach reproduces the standard results. But it also very naturally overcomes the
limitations of the usual treatment of systems of
identical particles. This GNS approach seems very general and can be applied for example to systems obeying
para- and braid- statistics including anyons.
\end{abstract}
\pacs{}
\keywords{}
\maketitle
\section{Introduction}
\label{sec:1}
In spite of the numerous efforts  to achieve a satisfactory understanding of  entanglement for systems of identical
particles, there is no  general agreement on the appropriate generalization of concepts valid for
non-identical constituents~\cite{Tichy2011}.
That is because many concepts  are usually only discussed in the context of quantum  systems  for which the
Hilbert space $\mathcal H$ is a simple tensor product with no additional structure.
An example is the Hilbert space $\mathcal H=\mathcal H_A  \otimes\mathcal H_B$ of two non identical particles.
In this
case the partial trace $\rho_A=\Tr_B |\psi\rangle\langle \psi|$ for $|\psi\rangle \in \mathcal H$ to obtain
the reduced density matrix has a good physical meaning: it corresponds to observations only on the subsystem
$A$.

In contrast, the Hilbert space of a system of $N$ identical bosons (fermions) is given by the symmetric (antisymmetric) $N$-fold tensor product of the single-particle spaces. The consequence is that any multi-particle state
contains \emph{intrinsic} correlations between subsystems due to quantum indistinguishability.
This, in turn, forces a departure from the straightforward application of entanglement-related concepts like
singular value decomposition (SVD), Schmidt rank, or  entanglement entropy.

In this context, Schliemann et al.~\cite{Schliemann2001} have introduced an analogue of the Schmidt rank, the `Slater rank'
to study entanglement in two-fermion systems, using a new version  of the SVD adapted to deal with
antisymmetric matrices. The extension of these ideas to the boson case was worked out in
\cite{Paskauskas2001} and \cite{Li2001}. These approaches also have not found general acceptance.

The problems arising in the interpretation of the Slater rank and the von Neumann entropy of the reduced
density matrix (obtained by partial tracing) for these systems have been analyzed in \cite{Ghirardi2002,Ghirardi2004}.

Numerous other proposals for the treatment of identical particles
have recently been put forward. But, as a closer look at the literature on the
subject~\cite{Adhikari2009,Amico2008,Banuls2009,Benatti2012,Cavalcanti2007,Eckert2002,
Ghirardi2002,Ghirardi2004,Horodecki2009,L'evay2005,Li2001,Paskauskas2001,Sasaki2011,
Schliemann2001,Shi2003,Tichy2011,Wiseman2003,Zanardi2002,Zander2012,Cavalcanti2007, Grabowski2011}
reveals, it is apparent that there is no consensus yet as to what the proper formalism should be.

In this paper, we propose an approach to the study of entanglement based on the theory of operator algebras.
The foundational results  of Gelfand,  Naimark and Segal on the representation theory of $C^*$-algebras,
abbreviated as the
\emph{GNS-construction}\cite{Gelfand1943,Segal1947,Haag1996} are used in order to obtain a generalized
notion of entanglement. In particular, the notion of \emph{partial trace} is replaced by the much more
general notion of \emph{restriction of a state to a subalgebra}\cite{Barnum2004}. This will allow us to treat entanglement of
identical and non-identical particles on an equal footing, without the need to resort to different separability
criteria according to the case under study.

In order to display the usefulness of our approach, several explicit examples will be worked out. In particular
we show that the GNS-construction gives zero for the von Neumann entropy of a fermionic or a bosonic state containing the
least possible amount of correlations. We believe that this settles an issue that has caused a lot of
confusion regarding the use of von Neumann entropy as a measure of entanglement for identical particles
\cite{Paskauskas2001,Wiseman2003,Ghirardi2004,Amico2008,Plastino2009}.
\section{The Basic Idea}
\subsection{Preliminary Remarks}
A vector state of a quantum system is usually  described by
a vector $|\psi\rangle$ in a Hilbert space $\mathcal H$
(pure case). More generally, a state is a density matrix $\rho:\mathcal H \rightarrow \mathcal H$, a linear
map satisfying  $\Tr \rho =1$ (normalization), $\rho^\dagger=\rho$ (self-adjointness) and positivity $\rho \geq 0$.
For \emph{pure} states the additional condition  $\rho^2=\rho$ is required, which amounts to the assertion
that $\rho$ is of the form $|\psi\rangle\langle\psi|$ for some normalized vector in $\mathcal H$.

Now, given that the expectation value of an observable $\mathcal O$ is defined by $\langle \mathcal O\rangle= \Tr (\rho \,\mathcal O)$, we can equivalently regard $\rho$ as a  \emph{linear functional} $\omega_\rho: \mathcal A\rightarrow \mathds C$
on a unital ($C^*$-) algebra $\mathcal A$ of observables
with unity $\mathds 1_A$ (we consider only unital algebras). That is,  $\omega_\rho(\mathcal O)\in \mathds C$, for $\mathcal O\in \mathcal A$. The normalization and positivity conditions above then take the form $\|\omega_\rho\|:=\omega_\rho(\mathds 1_A)=1$
and $\omega_\rho(\mathcal O^\dagger \mathcal O)\geq 0$ (for any $\mathcal O\in \mathcal A$).
Such a  positive linear functional of unit norm is called a  \emph{state} on the algebra $\mathcal A$.

As already mentioned, in the bipartite case $\mathcal H=\mathcal H_A\otimes \mathcal H_B$, the definition of
$\rho_A$ involves a partial trace operation. It is therefore natural to ask for a characterization of
this operation in  terms of the notion of states on an algebra.
For this purpose consider the subalgebra $\mathcal A_0$ consisting of all operators of the form
$K\otimes \mathds{1}_B$, for $K$ an observable on $\mathcal H_A$.
Defining a new state $\omega_0: \mathcal A_0\rightarrow \mathds C$ as the \emph{restriction} of
$\omega_\rho$ to the subalgebra $\mathcal A_0$, one easily checks that for any observable $K$ of $\mathcal H_A$
the equality $\omega_0(K\otimes \mathds 1_B)\equiv\Tr_A(\rho_A\,K)$ holds.

As we show below,  an algebraic description of the  quantum system where the basic objects are a
$C^*$-algebra  $\mathcal A$ and a state $\omega$  on the algebra
provides the solution to some of the problems that appear when $\mathcal H$ does not have
the form of a `simple tensor product'.

\subsection{The GNS Construction}

The basic idea of the GNS-construction is that given an algebra
$\mathcal A$ of observables and a state $\omega$ on this algebra, we can
construct the Hilbert space on which the algebra of observables acts on. The key steps are:
(a) Using $\omega$, we can endow $\mathcal A$ itself
with an inner product. So it becomes an `inner product' space  $\hat{\mathcal A}$.
(b) This inner product may be degenerate in the sense that the norm of some
non-null elements of $\hat{\mathcal A}$ may be zero. (c) If we remove these null vectors  by taking the
quotient of $\hat{\mathcal A}$ by the null space $\mathcal N$ of zero norm vectors to get
$\hat{\mathcal A}/\mathcal N$, then we have a well-defined positive
definite inner product on $\hat{\mathcal A}/\mathcal N$.
Hence we get a well-defined Hilbert space (after completion).
(d) The algebra of observables $\mathcal A$ acts naturally on this Hilbert space in a simple manner.

We now make this set of ideas more precise.

From the mathematical point of view, the algebra of observables is a
$C^*$-algebra. This guarantees that one has enough
structure to perform all the tasks in the list above.

A $C^*$-algebra is a (complete normed) algebra $(\mathcal A,\|\cdot\|)$,
together with an antilinear \emph{involution}
$\alpha\mapsto \alpha^*$, such that the basic property $\|\alpha^* \alpha\|=\|\alpha\|^2$ is satisfied
for all $\alpha$ in $\mathcal A$.
The prototypical  example of a $C^*$-algebra is the algebra $\mathcal B(\mathcal H)$ of all bounded operators
on a Hilbert space $\mathcal H$, with the involution given by the adjoint: $\alpha^*=\alpha^\dagger$.
Here we will only be interested in \emph{unital} algebras, that is, we assume the existence of a unit
$\mathds 1_{\mathcal A}$ for the algebra.

Given a state $\omega$ on a $C^*$-algebra $\mathcal A$, we can obtain a representation
$\pi_\omega$ of $\mathcal A$ on a Hilbert space $\mathcal H_\omega$ as follows.
Since $\mathcal A$ is an algebra, it is in particular a vector space.
When elements $\alpha \in \mathcal A$ are regarded as elements of a vector space $\hat{\mathcal A}$ we write them as
$|\alpha \rangle$. We then set $\langle \beta | \alpha \rangle~=~\omega(\beta^*\alpha)$. This is almost a
scalar product, $\langle \alpha | \alpha \rangle \geq 0$, but there could be a null space $\mathcal N_\omega$ of
zero norm vectors: $\mathcal N_\omega=\lbrace \alpha \in \mathcal  A \, |\, \omega (\alpha^* \alpha)=0\rbrace.$
Schwarz inequality shows that $\mathcal N_\omega$ is a left ideal:
\begin{equation}
a~\mathcal N_\omega~\in \mathcal N_\omega, \qquad \forall a \in\mathcal A.
\label{leftideal}
\end{equation}
It also shows that
\begin{equation}
\langle a|\alpha \rangle ~=~0, ~~\forall a \in \mathcal A, \alpha \in \mathcal N_\omega.
\end{equation}

We denote the elements of the quotient space
$\mathcal H_\omega = \hat{\mathcal A}/ \mathcal N_\omega$ by $|[\alpha]\rangle$, where $[\alpha]~=~\alpha +\mathcal N_\omega$
 $\forall\alpha \in\mathcal A$. It has a well-defined scalar product
\begin{equation}
\label{eq:inner}
\langle[\alpha] |[\beta]\rangle = \omega (\alpha^* \beta )
\end{equation}
(it is independent of the choice of $\alpha$ from $[\alpha]$ because of (\ref{leftideal}))
and no nontrivial null vectors. Moreover we have a representation
$\pi_\omega$ of $\mathcal A$ on $\mathcal H_\omega: \pi_\omega(\alpha)|[\beta]\rangle= |[\alpha\beta]\rangle$.
(To show this, use $\alpha \,\mathcal N_\omega \in \mathcal N_\omega$).

This representation $\pi_\omega$ is in general reducible. We decompose $\mathcal H_\omega$ into a direct sum of
irreducible spaces: $\mathcal H_\omega=\bigoplus_i\mathcal H_i$. Let
$P_i:\mathcal H_\omega\rightarrow\mathcal H_i$
be the corresponding orthogonal projections and define
\begin{equation}
\label{eq:mu-chi}
\mu_i=\|P_i\,|[\mathds 1_{\mathcal A}]\rangle\|\;\;(\mu_i>0\; \mbox{always})\; \;\;\;\;\;\mbox {and}\;\;\;\;\;\;
|[\chi_i]\rangle =(1/\mu_i) P_i|[\mathds 1_{\mathcal A}]\rangle.
\end{equation}
Observe that
\begin{equation}
\langle[\chi_i]|[\chi_j]\rangle ~=~\delta_{ij},
\end{equation}
and also  that
\begin{equation}
\omega(\alpha)~=~ \langle [\mathds 1_A]|\pi_\omega(\alpha) | [\mathds 1_A]\rangle,
~~~~|[\mathds 1_A]\rangle ~=~ \sum_i P_i |[\mathds 1_A]\rangle.
\end{equation}
One then obtains
\begin{equation}
\omega(\alpha)=\Tr_{\mathcal H_\omega}\left(\rho_\omega\,\pi_\omega(\alpha)\right),
\end{equation}
where $\rho_\omega$ is a density matrix on $\mathcal H_\omega$, given by
\begin{equation}
\rho_\omega = \sum_i\mu_i^2|[\chi_i]\rangle\langle [\chi_i]|.
\end{equation}
This follows from
\begin{equation}
\sum_{ij} \mu_j\mu_i\langle[\chi_i]|\pi_\omega(\alpha)|[\chi_j]\rangle~=~
\sum_i \mu_i^ 2\langle[\chi_i]|\pi_\omega(\alpha)|[\chi_i]\rangle,
\end{equation}
as $\pi_\omega(\alpha)$ has zero matrix elements between different irreducible subspaces.

This  is a crucial fact. Since $|[\chi_i]\rangle \langle [\chi_i]|$
is a pure state, it shows that $\omega$ is \emph{pure} if and only if the representation
$\pi_\omega:\mathcal A\rightarrow \mathcal B(\mathcal H_\omega)$ is \emph{irreducible},
a well-known result.
In particular, the von Neumann entropy of $\omega$,  $S(\omega)=-\Tr_{\mathcal H_\omega}\rho_\omega \log\rho_\omega$, is zero if and only if $\mathcal H_\omega$ is irreducible.
{\it{The latter is a property that depends on \emph{both} the algebra $\mathcal A$ and the state $\omega$}}.

Consider  now a (unital) subalgebra $\mathcal A_0\subset \mathcal A$ of  $\mathcal A$ and let
$\omega_0$ denote the \emph{restriction} to $\mathcal A_0$ of a pure state $\omega$ on
 $\mathcal A$
 ~\cite{*[{Importance of focusing on subsystems has also been emphasized by }] [{}] Barnum2004}.
We can apply the GNS-construction to the pair $(\mathcal A_0, \omega_0)$ and use the von Neumann entropy
of $\omega_0$ to study the entropy which arises from the restriction.
\subsection{Example 1: $M_2(\mathds C)$}
In order to illustrate the above GNS-construction, consider the algebra $\mathcal A = M_2(\mathds C)$
of $2\times2$ complex matrices.
Denoting by $e_{ij}$ the $2\times 2$ matrix with one on its $(i,j)$ entry and zero elsewhere,
we can write any $\alpha\in \mathcal A$ as $\alpha= \sum_{i,j\in\{1,2\}} \alpha_{ij}\, e_{ij}$.

It is readily checked that the map $\omega_\lambda: \mathcal A  \rightarrow \mathds C$ given by
$\omega_\lambda(\alpha)=\lambda \alpha_{11} + (1-\lambda)\alpha_{22}$
defines a \emph{state} on the algebra, as long as $0\leq \lambda \leq 1$.
The vector space $\mathcal{ \hat A}$ is  generated by  the four vectors
$|e_{ij}\rangle$ ($i,j=1,2$) and the null space $\mathcal N_{\omega_\lambda}$
is generated by those $\alpha$ such that $\omega_\lambda(\alpha^* \alpha)=0$.
Since
\begin{equation}
 \alpha^*\alpha ~=~ \sum_{ijk} \bar{\alpha}_{ki} \alpha_{kj} |i\rangle\langle j|,
\end{equation}
the explicit form of the null vector condition is:
 \begin{equation}
 \label{eq:condition}
 \lambda (|\alpha_{11}|^2 + |\alpha_{21}|^2)  +(1-\lambda) (|\alpha_{12}|^2 + |\alpha_{22}|^2)=0.
 \end{equation}
\noindent{\it{Case 1: $0 < \lambda < 1$}}

For $0<\lambda <1$ the only solution to this equation is $\alpha=0$, implying that there are no null vectors:
$\mathcal N_{\omega_\lambda}=\{0\}$. Therefore, in this case the GNS-space is given by
$\mathcal H_{\omega_\lambda}\cong\mathds C^4$.  The matrices $e_{ij}$ act on this space as :
\begin{equation}
\pi_{\omega_\lambda}(e_{ij})|[e_{kl}]\rangle= \delta_{jk}|[e_{il}]\rangle.
\end{equation}
The matrix  of $\pi_{{\omega_\lambda}}(e_{11})$ for instance is:
\begin{equation}
 \pi_{\omega_\lambda}(e_{11})=\left(
 \begin{array}{cccc}
 1 & 0 & 0 & 0\\
 0 & 1 & 0 & 0\\
 0 & 0 & 0 & 0\\
 0 & 0 & 0 & 0\\
 \end{array}
 \right),
 \end{equation}
 when we order the basis as $e_{11},e_{12},e_{21},e_{22}$.

This representation is clearly reducible, the subspaces $\mathcal H^{(l)}\; (l = 1,2)$ spanned by
$\{|[e_{kl}]\rangle\}_{k= 1,2}$ being invariant. Hence $\pi_\omega$ is the direct sum of two isomorphic irreducible
representations, the corresponding decomposition of $\mathcal H_{\omega_\lambda}$ being
\begin{equation}
\mathcal H_{\omega_\lambda}~=~ \mathcal H^{(1)} \oplus \mathcal H^{(2)}.
\end{equation}
Next we decompose $\omega_\lambda$ into pure states. The unity $\mathds 1_{\mathcal A}$ of $\mathcal A$ is just
$\mathds 1_2$ so that
\begin{equation}
|[\mathds 1_{\mathcal A}] \rangle~=|[e_{11}]\rangle + |[e_{22}]\rangle
\end{equation}
gives the decomposition of $|[\mathds 1_{\mathcal A}\rangle$ into irreducible subspaces.
The norms of the two components are $\sqrt{\lambda}$ and $\sqrt{1-\lambda}$ by (\ref{eq:inner}). So
\begin{equation}
|[\mathds 1_{\mathcal A}]\rangle~=~\sqrt{\lambda}|[\chi_{1}]\rangle + \sqrt{1-\lambda}|[\chi_{2}]\rangle,
\;\;\;\;\langle[\chi_{i}]|\chi_{j}]\rangle = \delta_{ij} ,
\end{equation}
with $|[\chi_i]\rangle$ as in (\ref{eq:mu-chi}). It follows that
\begin{equation}
\rho_{\omega_\lambda}~=~ \lambda |[\chi_{1}]\rangle \langle [\chi_{1}]| +
(1-\lambda) |[\chi_{2}]\rangle \langle [\chi_{2}]|,
\end{equation}
so that $\omega_\lambda$  is not pure. It has von Neumann  entropy
\begin{equation}
S(\omega_\lambda) = -\lambda\log \lambda - (1-\lambda)\log(1-\lambda).
\end{equation}
\noindent{\it{Case 2:  $\lambda ~=~0~or~1$}}

If we choose $\lambda=0$,
from (\ref{eq:condition}) we see that $\mathcal N_{\omega_\lambda}\cong \mathds C^2$, since it is spanned by elements of the form
 \begin{equation}
 \alpha=\left(\begin{array}{cc}
  \alpha_{11} & 0 \\
  \alpha_{21} & 0
 \end{array}\right),
 \end{equation}
 that is, by linear combinations of $|e_{11}\rangle$ and $|e_{21}\rangle$. Accordingly, the  GNS-space
$\mathcal H_{\omega_\lambda}= \hat{\mathcal A}/ \mathcal N_{\omega_\lambda}\cong \mathds C^2$ is generated by  $|[e_{12}]\rangle$ and $|[e_{22}]\rangle$.
In this case the representation of $\mathcal A$ is irreducible and given by $2\times 2$ matrices
$\pi_{\omega_\lambda}(e_{ij})$:
\begin{equation}
\pi_{\omega_\lambda}(e_{ij})|[e_{k2}]\rangle= \delta_{jk}|[e_{i2}]\rangle.
\end{equation}
The state $\omega_\lambda$ is pure with zero entropy. A similar situation is found for $\lambda=1$.
\subsection{Example 2: Bell State}
Let $\mathcal H=\mathcal H_A\otimes\mathcal H_B\equiv\mathds C^2\otimes \mathds C^2$ and consider the state
vector
\begin{equation}
|\psi\rangle=\frac{1}{\sqrt 2} \big(|+\rangle \otimes |-\rangle - |-\rangle\otimes|+\rangle\big).
\end{equation}
 Then
$|\psi\rangle \langle\psi|$
can be thought  of as a state $\omega$ on the algebra $\mathcal A$ of linear operators on $\mathcal H$.
This algebra  is isomorphic to the $4\times 4$ matrix algebra $M_4(\mathds C)$ and is generated by elements of the form
$\sigma_\mu\otimes\sigma_\nu$ ($\mu,\nu=0,1,2,3$), with $\sigma_0=\mathds 1_2$ and
$\{\sigma_{1},\sigma_{2},\sigma_{3}\}$ the Pauli matrices.

In this context, the  \emph{entanglement} of $|\psi\rangle$ is understood in terms of correlations between ``local"
measurements performed separately on subsystems $A$ and $B$. Measurements performed on $A$  correspond to
the restriction of $\omega$ to the subalgebra $\mathcal A_A\subset \mathcal A$ generated by elements of the
form $\sigma_\mu\otimes \mathds 1_2$.

We now study this case. We set $\omega_A= \omega\mid_{\mathcal A_A}$.
In order to construct the GNS-space $\mathcal H_{\omega_A}$, we first notice that
$\omega_A((\sigma_\mu\otimes \mathds 1_2)^*(\sigma_\nu\otimes \mathds 1_2))= \omega(\sigma_\mu^*\sigma_\nu\otimes
\mathds 1_2)=\langle \psi|\sigma_\mu^*\sigma_\nu\otimes \mathds 1_2|\psi\rangle=\delta_{\mu\nu}$,
so that in this case there are no nontrivial null states, that is,  $ \mathcal N_{\omega_A}= \{0\}$.
We obtain $\mathcal H_{\omega_A}=\mathcal{\hat A}_A/\mathcal N_{\omega_A}\cong \mathds C^4$, with basis
vectors $|[\sigma_\nu]\rangle \equiv |[\sigma_\nu\otimes \mathds 1_2]\rangle$ and inner product
$\langle[\sigma_\mu]|[\sigma_\nu]\rangle= \delta_{\mu\nu}$.

The action of  $\mathcal A_A$ on $\mathcal H_{\omega_A}$ is given, as explained above, by
linear operators $\pi_{\omega_A}(\alpha)$:
\begin{equation}
\pi_{\omega_A}(\alpha)|[\beta]\rangle = |[\alpha \beta]\rangle.
\end{equation}
The RHS  can be explicitly computed using
the identity $\sigma_i \sigma_j = \delta_{ij}\mathds 1_2 +i\varepsilon_{ijk} \sigma_k$. One then finds that the GNS
space splits into the sum of two invariant subspaces:
\begin{equation}
\mathcal H_{\omega_A}= \mathds C^2\oplus \mathds C^2.
\end{equation}
They are spanned by
\begin{equation}
\Big\{ \big|[\sigma_+\otimes \mathds 1_2]\big\rangle, ~~\big|[(1/2)(1-\sigma_3)\otimes \mathds 1_2]\big\rangle\Big\}
\end{equation}
and
\begin{equation}
\Big\{\big|[(1/2)(1+\sigma_3)\otimes \mathds 1_2]\big\rangle, ~~\big|[\sigma_-\otimes \mathds 1_2]\big\rangle\Big\},
\end{equation}
where
\begin{equation}
\sigma_\pm~=~\sigma_1 \pm i\sigma_2.
\end{equation}
The corresponding projections are
\begin{equation}
P_i=\frac{1}{2} \pi_{\omega_A}( \mathds 1_{\mathcal A} +(-1)^i \sigma_3 \otimes \mathds 1_2),\,\, \mbox{with } i=1,2.
\end{equation}
We obtain
\begin{equation}
  \mu_i^2=\|P_i|[\mathds 1_{\mathcal A}]\rangle\|^2=
\frac{1}{2}\omega_A(\mathds 1_{\mathcal A}+(-1)^i\sigma_3\otimes \mathds 1_2)=\frac{1}{2},
  \end{equation}
\begin{equation}
|[\chi_i]\rangle=\frac{1}{\sqrt 2} \big(|[\sigma_0]\rangle+(-1)^i|[\sigma_3]\rangle\big),\;\;\;
\mbox{with }\,\,\,\langle[\chi_i]|[\chi_j]\rangle =\delta_{ij}.
\end{equation}
Thus, the representation of $\omega_A$ as a density matrix on the GNS-space $\mathcal H_{\omega_A}$ is:
\begin{equation}
\rho_{\omega_A} = \frac{1}{2}|[\chi_1]\rangle\langle [\chi_1]|+\frac{1}{2}|[\chi_2]\rangle\langle [\chi_2]|.
\end{equation}
The von Neumann entropy computed via the GNS-construction is therefore
\begin{equation}
S(\omega_A)= \log 2,
\end{equation}
reproducing the standard result, as expected.
\section{Systems of Identical Particles}
Let $\mathcal H^{(1)}=\mathds C^d$ be the Hilbert space of a one-particle system. The group $U(d)=\{g\}$
acts on $\mathds C^d$ by the representation $U^{(1)}$
and the algebra of observables is given by a $*$-representation of  the \emph{group algebra}  $\mathds CU(d)$
on $\mathcal H^{(1)}$.
Its elements are of the form
\begin{equation}
 \widehat \alpha = \int_{U(d)}d\mu(g)  \alpha(g) U^{(1)}(g),
 \label{matrixalgebra}
\end{equation}
where $\alpha$ is a complex function on  $U(d)$ and
$\mu$ the Haar measure~\cite{Balachandran2010}.

The elements $\hat{\alpha}$ span the matrix algebra $M_d(\mathds C)$. We can understand (\ref{matrixalgebra})
in terms of matrix elements
\begin{equation}
\widehat \alpha_{ij} = \int_{U(d)}d\mu(g)  \alpha(g) U^{(1)}(g)_{ij}
\end{equation}
which are just the integrals of the functions $\alpha(g) U^{(1)}(g)_{ij}$.

Consider now a fermionic  system with single-particle space  $\mathcal H^{(1)}$.
The Hilbert space of this system
is the Fock space $\mathcal F = \bigoplus_{k=0}^d \mathcal H^{(k)}$, where $\mathcal H^{(k)}=\Lambda^k\mathcal H^{(1)}$
is the space of   antisymmetric $k$-tensors in $\mathcal H^{(1)}$.
Let $\{  |e_1\rangle, |e_2\rangle, \ldots, |e_d\rangle \}$  denote an orthonormal basis for $\mathcal H^{(1)}$.
Then, the set $\{ |e_{i_1}\wedge\ldots \wedge e_{i_k}\rangle\}_{1\leq i_1 < \cdots <i_k\leq d}$ provides an orthonormal basis for $\mathcal H^{(k)}$.

We can alternatively consider the canonical anticommutation relations (CAR)
algebra $\{a_i, a_j^\dagger\}=\delta_{ij}$, and obtain all basis vectors by repeated
application of creation operators to the vacuum vector $|\Omega\rangle$:
\begin{equation}
|e_{i_1}\wedge\ldots \wedge e_{i_k}\rangle=a_{i_1}^\dagger\ldots a_{i_k}^\dagger|\Omega\rangle.
\end{equation}
A self-adjoint operator $A$  on $\mathcal H^{(1)}$ can be made to act on $\mathcal H^{(k)}$
in a way that preserves the antisymmetric character of the vectors by considering combinations of the form
\begin{equation}
A^{(k)}~:=~(A\otimes\mathds{1}_d\cdots\otimes\mathds{1}_d) +
(\mathds{1}_d\otimes A\otimes\mathds{1}_d\otimes\cdots \otimes\mathds{1}_d)
+\cdots+
(\mathds{1}_d\otimes\cdots\otimes\mathds{1}_d\otimes A).
\label{liealgebra}
\end{equation}

{\it{The map $A \longrightarrow A^{(k)}$ is a Lie algebra homomorphism}}.
We further comment on this important point below.

At the group level, we may consider exponentials of such operators, of the form
$e^{i A}.$ We then see that the operators of the form
\begin{equation}
\label{eq:alpha^k}
\widehat \alpha^{(k)}~=   \int_{U(d)}d\mu (g) \alpha(g) U^{(1)}(g)\otimes\cdots\otimes U^{(1)}(g),
\end{equation}
act properly on $\mathcal H^{(k)}$.

{\it{ The map $\widehat{\alpha} \longrightarrow \widehat{\alpha}^{(k)}$
is an isomorphism from $M_d(\mathcal C)$ into $M_{d^k}(\mathcal C)$}}. This is also important as
discussed below.

These constructions are most conveniently expressed in terms of a \emph{coproduct} $\Delta$\cite{Balachandran2010}. In fact,
an approach based on Hopf algebras (as explained in~\cite{Balachandran2010}), has the great advantage that
para- and braid-statistics can be \emph{automatically} included. In our present case, the construction of the
observable algebra corresponds to the following simple choice for the coproduct:
$\Delta (g) = g\otimes g$, ($g\in U(d)$), linearly extended to all of $\mathds C U(d)$.
This choice fixes the form of (\ref{eq:alpha^k}). At the Lie algebra level, it reduces to (\ref{liealgebra}).

Physically, the existence of such a coproduct is very important,
since it allows us to homomorphically  represent the one-particle observable algebra on the $k$-particle sector.
In a many particle system, the choice to perform observations of only one-particle observables corresponds to
restricting the full-algebra of observables to the homomorphic image of the one-particle observable algebra
obtained by using the coproduct.

Now, if for some reason we perform only partial one-particle observations
(for instance, if at the one-particle level, we decide to measure -or have access to-  only the spin degrees of
freedom, or only the position), the one-particle observable algebra will have to be restricted accordingly and
hence its homomorphic image at the $k$-particle level will be a subalgebra of the original algebra.

As we will see in what follows, the GNS approach covers all such cases whether or not particles are
identical. In particular it merits to reemphasize that it covers observations of
particles obeying para- and braid- statistics, including anyons.

\subsection{Example 3: Two Fermions, $\mathcal H^{(1)}=\mathds C^3$.}

Keeping the same notation as above,  put $d=3$. We focus our attention on the
two-fermion space $\mathcal H^{(2)}=\Lambda^2 \mathds C^3\subset\mathcal H^{(1)}\otimes \mathcal H^{(1)}$,
with basis
\begin{equation}
\label{eq:3-bar-basis}
\big\{|f^k\rangle:=\varepsilon^{ijk}|e_i \wedge e_j \rangle\big\}_{1\leq k\leq 3},
\end{equation}
 $\big\{|e_1\rangle, |e_2\rangle,|e_3\rangle\big\}$ being  an orthonormal basis for $\mathcal H^{(1)}$.
The algebra $\mathcal A$ of observables for the two-fermion system is the
matrix algebra generated by $|f^i\rangle\langle f^j|$ ($i,j=1,2,3$). It is isomorphic to $M_3(\mathds C)$.

Now,  $U(3)$ acts   on    $\mathcal H^{(1)}$ through the defining representation ($U^{(1)}(g)=g$), so that
one particle observables are given - at the  two fermion level -
by the action of $\mathds C U(3)$  on $\mathcal H^{(2)}$. This action is given
by the restriction of the operators
$\widehat \alpha=   \int_{U(d)}d\mu(g)  \alpha(g) U^{(1)}(g)\otimes U^{(1)}(g)$
to the space of antisymmetric vectors.
Let 3 be the defining (or fundamental) $SU(3)$ representation on $\mathcal H^{(1)}$. Then the restriction can be obtained from
the decomposition $3\otimes 3 = 6\oplus \bar 3$ of the $SU(3)$ representation. The
$|f^i\rangle$ span this $\bar 3$ representation.

\noindent{\it {Choice 1 for $\mathcal A_0$ :}}

Let $|\psi\rangle\in \mathcal H^{(2)}$ be any two-fermion state vector. If we take $\mathcal A_0$ to be the
full algebra of one-particle observables acting on $\mathcal H^{(2)}$,
then  $\mathcal A_0=\mathcal A$ and the GNS-representation
corresponding to the pair  $(\mathcal A_0, \omega_\psi)$ is irreducible, the state
remaining unchanged upon restriction.
This is just the fact that the $\bar{3}$ representation of $SU(3)$ is irreducible and
corresponds to the fact that, for $d=3$, all two-fermion vector
states $|\psi\rangle$ have Slater rank 1.

We hence get zero for the von Neumann entropy.

Notice, however, that the {\it{von Neumann entropy computed by
partial trace is equal to 1 for all choices of $|\psi\rangle$ (cf.~\cite{Ghirardi2004}),
in disagreement with the GNS-approach}}.

\noindent{\it {Choice 2 for $\mathcal A_0$ :}}

The situation changes drastically if  we make a different choice for the subalgebra $\mathcal A_0$ of $\mathcal A$.
Let us, for the sake of concreteness, choose $\mathcal A_0$ to be  given by those one-particle observables
pertaining \emph{only} to the one-particle states $|e_1\rangle$ and $|e_2\rangle$.
In this case,  $\mathcal A_0$ will be the five dimensional algebra generated by
$M^{ij}:=|f^i\rangle\langle f^j|$ $(i,j = 1,2)$  and $\mathds 1_{\mathcal A}$, the unit matrix.

As a physical illustration for the meaning of $M^{ij}$, let us think of  $e_1,e_2,e_3$  as $ u,d,s$ quarks. Observables acting just on $u$ and $d$ amounts to  the isospin group $SU(2)_I$ acting just on $|f^1\rangle$ and $|f^2\rangle$ in the $\bar{3}$
representation. And the group algebra  ${\mathds C}SU(2)_I$ is also generated by $M^{ij}$.

We now  explicitly perform the GNS-construction  for the particular choice
\begin{equation}
|\psi_\theta\rangle=   \cos\theta|f^1\rangle + \sin\theta |f^3\rangle.
\end{equation}
Let $\omega_\theta: \mathcal A  \rightarrow  \mathds C$ denote the corresponding state:
\begin{equation}
\omega_\theta (\alpha)= \langle\psi_\theta |\alpha|  \psi_\theta \rangle,\;\;\;\;\forall\alpha\in \mathcal A,
\end{equation}
and put
\begin{equation}
\omega_{\theta,0}=\omega_\theta\mid_{\mathcal A_0}.
\end{equation}
Now notice that $\omega_{\theta,0}(M^{12 *}M^{12})=0$ independently of $\theta$.
The same holds true for  $M^{22}$ so that both are null vectors:
$|[M^{12}]\rangle = |[M^{22}]\rangle= 0$.

\noindent{\it{Case 1: $0 < \theta < \frac{\pi}{2}$}}

For $0<\theta<\pi/2$, there are no more linearly independent null vectors.
We can see this from
\begin{equation}
 \langle\psi_\theta |\alpha^*\alpha | \psi_\theta \rangle = 0 \Rightarrow \alpha|\psi_\theta \rangle = 0
\Rightarrow \alpha = \sum c_iM^{i2}, c_i \in \mathds C.
\end{equation}
Therefore, the null space $ \mathcal N_{\theta,0}$ is two-dimensional and the GNS-space
$\mathcal H_\theta= \hat{\mathcal A_0}/\mathcal N_{\theta,0}$ is the three-dimensional space with basis
$\{|[M^{11}]\rangle, |[M^{21}]\rangle,|[E^3]\rangle\}$, where
$E^3:=\mathds 1_{\mathcal A}-M^{11}-M^{22}$.

Since $\alpha_0\, E^3=0$ if $\alpha_0 \in \mathcal A_0$,
we immediately recognize that, in terms of irreducibles,
$\mathcal H_\theta=\mathds C^2\oplus\mathds C^1$. Call $P_1$ and $P_2$ the corresponding projections.
After noting that $[M^{11}+ M^{22}] ~=~ [\mathds 1_2]$, we obtain
\begin{equation}
P_1 |[  \mathds 1_{\mathcal A} ]\rangle=|[ M^{11} ]\rangle ,~~
P_2 |[  \mathds 1_{\mathcal A} ]\rangle=|[ E^3  ]\rangle
\end{equation}
The corresponding `weights' $|\mu_i|^2= \|P_i |[  \mathds 1_{\mathcal A} ]\rangle\|^2$ are computed using the inner product of $\mathcal H_{\theta}$. We obtain
\begin{equation}
|\mu_1|^2 = \cos^2\theta~,~~|\mu_2|^2 = \sin^2\theta.
\end{equation}
Hence,
\begin{equation}
\omega_{\theta,0}~=~\cos^2\theta\left(\frac{1}{\cos^2\theta} |[M^{11}]\rangle \langle[M^{11}]|\right)
+ \sin^2\theta\left(\frac{1}{\sin^2\theta} |[E^3]\rangle \langle[E^3]|\right).
\end{equation}
The result for the  entropy as a function of $\theta$ is therefore
\begin{equation}
S(\theta) = -\cos^2 \theta \log \cos^2\theta - \sin^2 \theta \log \sin^2\theta.
\end{equation}

\noindent{\it {Case 2: $\theta = 0$}}

It is readily checked that at $\theta=0$  additional null states appear
as compared to Case 1.
Thus, for $\theta = 0$ the null vectors  are spanned by
\begin{equation}
|M^{12}\rangle,\;\;\; |M^{22}\rangle,\;\;\; |E^3\rangle.
\end{equation}
The GNS space $\mathcal H_0~=~\hat{\mathcal A_0}/\mathcal N_{0,0}$ is two-dimensional
and irreducible. It is spanned by
\begin{equation}
 |[M^{11}]\rangle, \;\;\;|[M^{21}]\rangle.
\end{equation}
Since $\pi_\omega(\mathcal A_0)$ acts nontrivially on this space, and the smallest nontrivial representation
of $\mathcal A_0$ is its two-dimensional IRR, this representation is irreducible. Hence $\omega_{0,0}$
is pure with zero entropy.
For completeness we note that the projector to $\mathcal H_0$ is
$\pi_\omega(M^{11}+M^{22})$.
\newpage
\noindent{\it {Case 3: $\theta = \frac{\pi}{2}$}}

For $\theta = \frac{\pi}{2}$ instead, {\it{all}} of $|M^{ij}\rangle$ are null vectors. So
$\mathcal H_{\frac{\pi}{2}} $ is one-dimensional and spanned by $|[E^3]\rangle $. Clearly
$\omega_{\frac{\pi}{2}}$ is pure with zero entropy.

The decomposition of $\mathcal H_\theta$ as a direct sum of irreducible subspaces
as we change the value of $\theta$ is, therefore,  as follows:
\begin{eqnarray}
\mathcal H_\theta\cong \left\lbrace\begin{array}{cc}
\mathds C^2,&\;\theta=0\\
\mathds C^3\cong \mathds C^2\oplus \mathds C, &\; \theta\in (0,\pi/2)\\
\mathds C,\; &\theta=\pi/2.
\end{array}
\right.
\end{eqnarray}
This result should be contrasted against the fact that the $\bar 3$ representation,
when regarded as a representation space for  $SU(2)$ acting on $|[M^{11}]\rangle$, $|[M^{22}]\rangle$
and $|[E^{3}]\rangle$
splits as $2\oplus 1$.

\subsection{Example 4: Two Fermions, $\mathcal H^{(1)}=\mathds C^4$}
Consider, in the  spirit of \cite{Eckert2002},  a one-particle space describing fermions with two
degrees of freedom which we call \emph{external} (e.g. \emph{`left'} and \emph{`right'}) and two
degrees of freedom which we call \emph{internal}
e.g. \emph{`spin 1/2'}). Here it is convenient to use a description in terms of
fermionic creation/annihilation operators
$a^{(\dagger)}_\sigma, b^{(\dagger)}_\sigma$, with $a$ standing for \emph{`left'},
$b$ for  \emph{`right'} and $\sigma=1,2$ for  spin up and down, respectively.
A basis for $\mathcal H^{(2)}$ is then given by the vectors $a^\dagger_1 a^\dagger_2|\Omega\rangle$,
$b^\dagger_1 b^\dagger_2|\Omega\rangle$ and $a^\dagger_\sigma b^\dagger_{\sigma'}|\Omega\rangle$, with
$\sigma,\sigma'\in\{ 1,2\}$.

Again we consider a $\theta$-dependent state vector, this time given by
\begin{equation}
|\psi_\theta\rangle=\left(\cos\theta a_1^\dagger b_2^\dagger + \sin\theta a_2^\dagger b_1^\dagger\right)
|\Omega\rangle.
\label{thetastate}
\end{equation}
At the two-particle level, the full observable algebra $\mathcal A$ is the matrix algebra
$M_{6}(\mathds C)$. From this algebra we pick the subalgebra $\mathcal A_0$ of
%Comment AR: changed $M_{36}(\mathds C)$ by $M_{6}(\mathds C)$..dim = 36 , but they are 6 x 6 matrices
\emph{one-particle observables corresponding to measurements at the left  location}.
This is the six-dimensional algebra generated by
\begin{eqnarray}
&\mathds 1_{\mathcal A},\; \;
T_1:=\frac{1}{2}(a_1^\dagger a_2 + a_2^\dagger a_1 ),\;\;
T_2:=-\frac{i}{2}(a_1^\dagger a_2 - a_2^\dagger a_1 ),\;\;
T_3:=\frac{1}{2}(a_1^\dagger a_1 - a_2^\dagger a_2 ),&\nonumber\\
&n_{12}:= (a_1^\dagger a_1 a_2^\dagger a_2),\;\;
N_a:=(a_1^\dagger a_1 + a_2^\dagger a_2 ).&
\end{eqnarray}
\noindent{\it{Case 1: $0 < \theta < \frac{\pi}{2}$}}

For $0<\theta < \pi/2$ we readily find that a basis of
null vectors of $\omega_\theta~=~|\psi_\theta \rangle \langle \psi_\theta|$
are $|n_{12}\rangle $ and $|(\mathds 1_{\mathcal A}-N_a)\rangle $. The GNS Hilbert space $\mathcal H_\theta =
{\hat{\mathcal A_0}}/{\mathcal N_{\theta,0}}$ is
hence four-dimensional and
spanned by the  vectors $|[\mathds 1_{\mathcal A}]\rangle$ and $\{|[T_i]\rangle\}_{i=1,2,3}$.

Let $\pi_\theta$ be the GNS representation of $\mathcal A_0$ on $\mathcal H_\theta$.
We can evidently find the decomposition of $\mathcal H_\theta$
into irreducible subspaces under $\pi_\theta$ by computing the
Casimir operator and the highest weight vectors of the Lie algebra $\mathfrak{su}(2)$ given by the representation
$T_i\mapsto  \pi_\theta(T_i)$.
We find  $\mathcal H_\theta=\mathcal H_1\oplus\mathcal H_2$, with $\mathcal H_1$
spanned by $|[T_1+iT_2]\rangle=|[a_1^\dagger a_2]\rangle$ and $|[a_2^\dagger a_2]\rangle$
and $\mathcal H_2$ spanned by $|[a_1^\dagger a_1]\rangle$ and $|[T_1-iT_2]\rangle=|[a_2^\dagger a_1]\rangle$.
The two representations are isomorphic.

We can find the components of $|[\mathds 1_{\mathcal A}]\rangle$ into $\mathcal H_i$ by
writing
\begin{equation}
|[\mathds 1_{\mathcal A}]\rangle = |[N_a]\rangle = |[a_1^\dagger a_1 + a_2^\dagger a_2]\rangle.
\end{equation}
Hence
\begin{equation}
P_1|[\mathds 1_{\mathcal A}]\rangle= |[a_2^\dagger a_2]\rangle,~~
P_2|[\mathds 1_{\mathcal A}]\rangle= |[a_1^\dagger a_1]\rangle.
\end{equation}
As for their normalization, using (\ref{thetastate}),
\begin{equation}
\|P_1|[\mathds 1_{\mathcal A}]\rangle\|^2 = \sin^2\theta,~~~~~
\|P_2|[\mathds 1_{\mathcal A}]\rangle\|^2 = \cos^2\theta.
\end{equation}
Hence the restriction $\omega_{\theta,0}$ of $\omega_\theta = |\psi_\theta\rangle
\langle \psi_\theta|$ to $\mathcal A_0$ can be written in terms of pure states as
\begin{equation}
\omega_{\theta,0} = \sin^2\theta |\chi_1\rangle \langle \chi_1| + \cos^2\theta|\chi_2\rangle \langle \chi_2|, \;\;\;\mbox{with}
\end{equation}
\begin{equation}
|\chi_1\rangle = \frac{1}{\sin\theta} |[a_2^\dagger a_2]\rangle,\;\;\;
|\chi_2\rangle = \frac{1}{\cos\theta} |[a_1^\dagger a_1]\rangle,\;\;\;
 \langle\chi_i|\chi_j\rangle = \delta_{ij}.
\end{equation}
This gives the following result for entropy:
\begin{equation}
S(\theta) = -\cos^2 \theta \log \cos^2\theta -\sin^2 \theta \log \sin^2\theta.
\end{equation}

\noindent{\it {Case 2: $\theta=0,\frac{\pi}{2}$}}

Consider first $\theta = 0$. In this case,
\begin{equation}
|\psi_0\rangle = a_1^\dagger b_2^\dagger |\Omega \rangle
\end{equation}
The null vectors $|\alpha\rangle$ are obtained by solving $\alpha |\psi_0\rangle = 0$ for $\alpha \in
\mathcal A_0$. That shows that
\begin{equation}
\mathcal N_{0,0} = \mbox{Span}\left\{|n_{12}\rangle, |[{\mathds 1_{\mathcal A}} - a_1^\dagger a_1] \rangle,
|[a_2^\dagger a_2]\rangle, |[a_1^\dagger a_2]\rangle\right\}.
\label{null}
\end{equation}
The quotient space ${\hat{\mathcal A_0}}/{\mathcal N_{0,0}}$ is $\mathds C^2$ and is isomorphic to
$\mathcal H_2$ above:
\begin{equation}
{\hat{\mathcal A_0}}/{\mathcal N_{0,0}}~
=\mbox{Span}\left\{|[a_1^\dagger a_1] \rangle, |[a_2^\dagger a_1]\rangle\right\}
= \mathds C^2.
\label{quotient}
\end{equation}
For $\theta = \frac{\pi}{2}$, when $|\psi_{\pi/2}\rangle~=~a_2^\dagger b_1^\dagger |0\rangle $, we find
instead a $\mathds C^2$ isomorphic to $\mathcal H_1$ above:
\begin{eqnarray}
\mathcal N_{\frac{\pi}{2},0} &=& \mbox{Span} \left\{ |[n_{12}]\rangle, |[\mathds 1_{\mathcal A}-
a_2^\dagger a_2]\rangle, |[a_1^\dagger a_1]\rangle, |[a_2^\dagger a_1]\rangle \right\}, \\
{\hat{\mathcal A_0}}/{\mathcal N_{\frac{\pi}{2},0}} &=& \mbox{Span} \left\{ |[a_1^\dagger a_2]\rangle, |[a_2^\dagger a_2]
\rangle\right\} = \mathds C^2
\end{eqnarray}
The GNS representations on both these $\mathds C^2$'s are irreducible.
Hence $\omega_{0,0}$ and $\omega_{\frac{\pi}{2},0}$
are pure states with zero entropy.

The decomposition of $\mathcal H_\theta$ into irreducible subspaces,
as a function of $\theta$, is the following:
\begin{eqnarray}
\mathcal H_\theta\cong \left\lbrace\begin{array}{cc}
\mathds C^2,&\;\theta=0, \pi/2\\
\mathds C^4\cong \mathds C^2\oplus \mathds C^2, &\; \theta\in (0,\pi/2).\\
\end{array}
\right.
\end{eqnarray}
The significant aspect of this example is the fact that for the values of $\theta $ for which the
Slater rank of $|\psi_\theta\rangle$ is one, namely $\theta = 0$ and $\frac{\pi}{2}$,
we obtain exactly zero for the entropy. In previous
treatments of entanglement for identical particles, the minimum value for the von Neumann entropy
of the reduced density matrix (obtained by partial trace) has been found to be $\log 2$ (cf. \cite{Tichy2011} and references therein). This has been
a source of embarrassment: it seems to suggest that different entanglement criteria have to be adopted,
depending on whether one is dealing with non-identical particles, or with bosons, or fermions.

We have shown here  that, by replacing the notion of \emph{partial trace} by the more general
one of \emph{restriction to a subalgebra}, all cases can be treated on an equal footing.

\subsection{Example 5: Two Bosons, $\mathcal H^{(1)}=\mathds C^3$.}
%%%%%%%%%%%%%%%%%%%%%%%%%%%%%%%%%%%%%%%%%%%%%%%%%%%%%%%%%%%%%%%%%%%%%%%%%%%%%%%%%%%%%%%%%%%%%%%%%%%%%%%%%%%%%%%%%%%
Here we consider the bosonic analogue  of Example 3. Consider the one-particle
space $\mathcal H^{(1)}=\mathds C^3$ with an orthonormal basis
$\{|e_1\rangle, |e_2\rangle,|e_3\rangle\}$. The two-boson space $\mathcal H^{(2)}$
is the space of symmetrized vectors in  $\mathcal H^{(1)}\otimes \mathcal H^{(1)}$. It corresponds to the six-dimensional space obtained from the decomposition $3\otimes 3 = 6\oplus \bar 3$ of $SU(3)$.
An orthonormal basis for $\mathcal H^{(2)}$ is given by vectors $\{|e_i \vee e_j\rangle \}_{i,j\in \{1,2,3\}}$, where
\begin{equation}
|e_i \vee e_j\rangle \equiv\left\{
\begin{array}{cc}
\frac{1}{\sqrt 2}(|e_i\rangle \otimes |e_j\rangle + |e_j\rangle \otimes |e_i\rangle), & i\neq j,\\
|e_i\rangle \otimes |e_i\rangle, &i =j.
\end{array}
\right.
\end{equation}
The algebra $\mathcal A$ of observables for the two boson system is thus isomorphic to $M_6(\mathds C)$.

Consider the state $\omega_{(\theta,\phi)}: \mathcal A\rightarrow \mathds C$ corresponding to the vector
\begin{equation}
\label{eq:bosons}
|\psi_{(\theta,\phi)}\rangle = \sin\theta\cos\phi|e_1 \vee e_2\rangle +\sin\theta\sin\phi|e_1 \vee e_3\rangle + \cos\theta|e_3 \vee e_3\rangle.
\end{equation}

We are interested in the restriction $\omega_{(\theta,\phi)}$ to the subalgebra $\mathcal A_0$ of one-particle observables
pertaining \emph{only} to the one-particle vectors $|e_1\rangle$ and $|e_2\rangle$.
Proceeding in the same way as in the previous examples, we recognize that the 6 representation,
when regarded as a representation space for $SU(2)$ acting nontrivially on
$|e_1\rangle$ and $|e_2\rangle$, splits as $6=3\oplus 2\oplus 1$. The basis vectors for these three
invariant subspaces are given below:
\begin{eqnarray}
3:~~&|1\rangle = |e_1 \vee e_1\rangle,\; \; |0\rangle = |e_1 \vee e_2\rangle, \; \; |-1\rangle = |e_2 \vee e_2\rangle, &
\nonumber\\
2:~~&|1/2\rangle = |e_1 \vee e_3\rangle,\;\; |-1/2\rangle=|e_2 \vee e_3\rangle, &\\
1:~~& |\tilde 0\rangle = |e_3 \vee e_3\rangle. &\nonumber
\label{invariant}
\end{eqnarray}
The one-particle observables on $\mathcal H^{(2)}$ are obtained
from  the operators $|e_i\rangle\langle e_j|$ (with $i,j=1,2$),
as well as from the unit operator on $\mathcal H^{(1)}$, by means of the coproduct.
Thus, the  subalgebra $\mathcal A_0$ is generated
by operators of the form $|u\rangle\langle v|$, with both $|u\rangle$ and $|v\rangle$ belonging to the \emph{same}
irreducible component of $\mathcal H^{(2)}$. (Note that the image of unity on $\mathcal H^{(1)}$ under the coproduct
$\Delta$ is $\mathds 1_{\mathcal A}$. Hence by taking combinations of images
of the above $\mathcal H^{(1)}$-observables under
$\Delta$, we see that $\mathcal A_0$ contains $|\tilde 0\rangle \langle \tilde 0|$).
In other words, $\mathcal A_0$ is given by block-diagonal matrices, with each block corresponding
to one of the irreducible components in the decomposition $6=3\oplus 2\oplus 1$. The dimension of
$\mathcal A_0$ is therefore $3^2+2^2+1^2 ~=~ 14$.

The construction of the GNS-representation corresponding to each particular value of the parameters
$\theta$ and $\phi$ is performed following the same procedure as in Example 3. Let us introduce the
notation $B_{u, v}\equiv |u\rangle\langle v|$, for any pair $|u\rangle, |v\rangle $ in (\ref{invariant}). Then,
from (\ref{eq:bosons}) we see that as long as the $(\theta,\phi)$-coefficients are all different from zero,
those elements of $\mathcal A_0$ of the form $B_{j,\pm 1}$ ($j=0,\pm 1$) and $B_{\sigma,-1/2}$ ($\sigma=\pm 1/2$)
generate the null vectors. That these generate all the null vectors
follows from the fact that (\ref{eq:bosons}) contains
one basis element for every irreducible component,
so that no further linear relation can arise that lead to null vectors.
So in this case we have
\begin{equation}
\mathcal H_{(\theta,\phi)} : = {\hat{\mathcal A_0}}/{\mathcal N_{(\theta,\phi),0}} = \mathds C^6,
~~~~\mathcal N_{(\theta,\phi),0} = {\rm Null~space~}.
\end{equation}
In terms of irreducible subspaces, one can readily see that $\mathds C^6$ decomposes according to
$\mathds C^6 = \mathds C^3 \oplus \mathds C^2 \oplus \mathds C^1$.

In general, we can read off the decomposition of $\mathcal H_{(\theta,\phi)}$ into irreducible subspaces
from (\ref{eq:bosons}), depending on which of its coefficents vanish.
For example, if only the first one vanishes,
\begin{equation}
\mathcal H_{(\theta,\phi)} = \mathds C^2 \oplus \mathds C^1.
\end{equation}
It is interesting to consider  the entropy as a function of $(\theta,\phi)$. For the case
in which all $(\theta,\phi)$-coefficients are non-zero,  we have:
\begin{equation}
|[\mathds 1_{\mathcal A}]\rangle= |[B_{1,1}]\rangle+ |[B_{1/2,1/2}]\rangle + |[B_{\tilde 0, \tilde 0}]\rangle,
\end{equation}
from which the entropy is readily computed as before. The result is:
\begin{equation}
\label{eq:boson-entropy}
S(\theta,\phi) = -\sin^2\theta[\cos^2\phi\log (\sin \theta \cos\phi)^2 +
  \sin^2\phi \log (\sin \theta \sin\phi)^2] -
 \cos^2\theta\log (\cos\theta)^2.
\end{equation}

The analytic formulae for entropy when one or more of the coefficients in (\ref{invariant})
vanish can be obtained from
(\ref{eq:boson-entropy}) by taking suitable limits on $\theta$ and $\phi$.

\begin{figure}
\includegraphics[scale=.9]{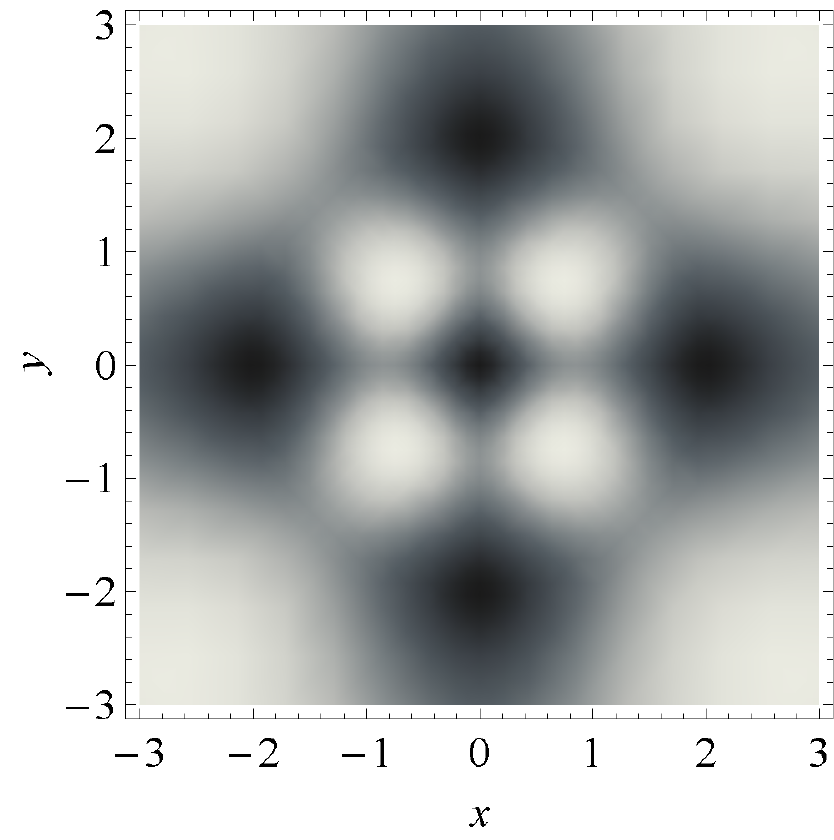}
\caption{\label{fig}
The entropy equation (\ref{eq:boson-entropy}) as a function of $x$ and $y$, the coordinates of a plane
representing the $(\theta,\phi)$-sphere through stereographic projection. Darker regions correspond to lower values of the entropy. Five of the six vanishing points of the entropy can be seen on the picture (black spots). The sixth one, corresponding to the north-pole of the sphere, lies `at infinity' in this representation.}
\end{figure}

We can see that the entropy vanishes whenever
$|\psi_{(\theta,\phi)}\rangle$ lies in a single irreducible component.
This happens precisely at those points of the two-sphere generated by the parameters $(\theta,\phi)$
that correspond to the coordinate axes. There are therefore six points where the entropy vanishes exactly.
This is depicted in Figure \ref{fig}, where the $(\theta,\phi)$-sphere has been mapped to the $x$-$y$ plane
through a stereographic projection. The figure shows the entropy as a function of the coordinates of
that plane.

%%%%%%%%%%%%%%%%%%%%%%%%%%%%%%%%%%%%%%%%%%%%%%%%%%%%%%%%%%%%%%%%%%%%%%%%%%%%%%%%%%%%%%%%%%%%%%%%%%%%%%%%%%%%%%%%%%
\section{Conclusions}
We have presented  a new approach to the study of quantum entanglement based on restriction of states to subalgebras.
The GNS-construction allows us to obtain a representation space for the subalgebra such that its decomposition
into irreducible subspaces can be used to study quantum correlations. We showed that, when applied
to bipartite systems for which the Hilbert space is a `simple' tensor product, our method reproduces the standard
results on entanglement. We furthermore showed, with explicit examples, how the formalism can be
applied to systems of identical particles. Our results demonstrate in a clear fashion that the von Neumann entropy
indeed remains a suitable entanglement measure, when understood in terms of states on algebras of observables.

The formalism used for the treatment of identical particles, using coproducts to identify algebras of
subsytems can be easily generalized to more sophisticated situations such as those of
particles obeying para- and braid statistics. Our results can hence be
extended to the study of entanglement of such particles. It can even be extended to study for instance
a $k$-particle subsystem in an $N$ particle Hilbert space.
\begin{acknowledgments}
\section{Acknowledgments}
The authors would like to thank Alonso Botero for  discussions that led to this work.
 ARQ and AFRL  acknowledge the warm hospitality of  T.R. Govindarajan at
The Institute of Mathematical Sciences, Chennai, where the main part of this work was done.
We also thank M. Asorey, S. Ghosh, K. Gupta, A. Ibort, G. Marmo and V. P. Nair for fruitful discussions.
APB is supported by DOE under grant number DE-FG02-85ER40231 and by the Institute of Mathematical Sciences,
Chennai. ARQ is supported by CNPq under process number 307760/2009-0.
AFRL is supported by Universidad de los Andes.
%acknowledges financial support by the Science Faculty of Universidad de los Andes.
 \end{acknowledgments}
%\bibliography{GNS}
%
\end{document}